# Magnetic Resonance Imaging of Single Atoms

Philip Willke[1,2,3], Kai Yang[1], Yujeong Bae[1,2,3], Andreas J. Heinrich[2,3,*] and Christopher P. Lutz[1,*]

[1] IBM Almaden Research Center, San Jose, CA 95120, USA
[2] Center for Quantum Nanoscience, Institute for Basic Science (IBS), Seoul 03760, Republic of Korea
[3] Department of Physics, Ewha Womans University, Seoul 03760, Republic of Korea

E-Mail: cplutz@us.ibm.com, heinrich.andreas@qns.science

**Magnetic resonance imaging (MRI) revolutionized diagnostic medicine and biomedical research by allowing a noninvasive access to spin ensembles[1]. To enhance MRI resolution to the nanometer scale, new approaches[2–4] including scanning probe methods[5–7] have been used in recent years, which culminated in detection of individual spins[5,6]. This allowed three-dimensional (3D) visualization of organic samples[8] and of sophisticated spin-structures[9-12]. Here, we demonstrate for the first time MRI of individual atoms on a surface. The setup, implemented in a cryogenic scanning tunneling microscope (STM), uses single-atom electron spin resonance (ESR)[13,14] to achieve sub-Ångström resolution exceeding the spatial resolution of previous experiments by one to two orders of magnitude. We find that MRI scans of different atomic species and probe tips lead to unique signatures in the resonance images. These signatures reveal the magnetic interactions between the tip and the atom, in particular magnetic dipolar and exchange interaction.**

The potential applications of atomic-scale MRI include imaging biomolecules[8] with unprecedented resolution, revealing the spin structure of atoms, molecules and solids[9-12], and giving site-dependent control in quantum simulators[15] and spin networks[16]. Scanning-field-gradient techniques have succeeded in sensing single electron spins[5-7] and nuclear spin ensembles[17-19], by employing magnetic resonance force microscopy (MRFM)[5] and scanning probes using nitrogen-vacancy (NV) centers[6]. In those studies, the field gradient of a magnetic tip was used to locally tune the resonance frequency of spin ensembles and individual spin centers. These techniques increased the resolution of MRI by several orders of magnitude, down to a few nanometers[5], in special cases to sub-nanometer resolution[20], and they can operate at room temperature[21]. However, the spatial resolution of these resonance techniques is still lower than that of conventional low-temperature scanning probe microscopy. Conversely, scanning probe methods have been used to map electrostatic[22] and magnetic interaction potentials[23], but their energy resolution is limited in tunneling spectroscopy experiments by thermal broadening[24].

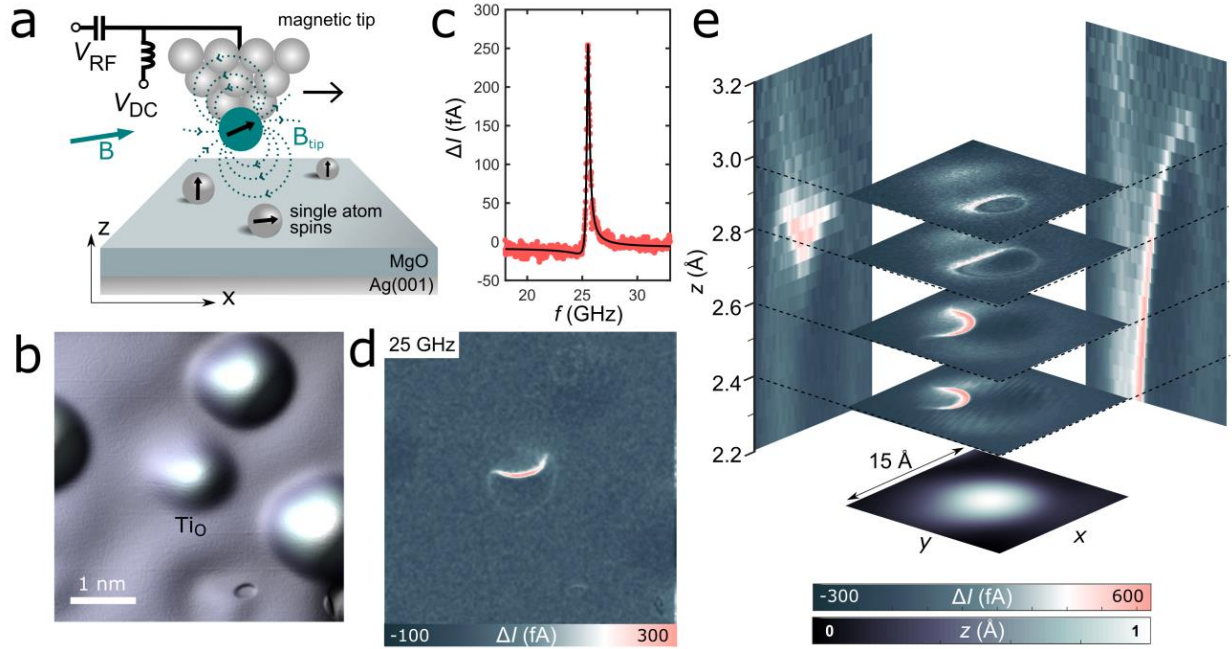

**Figure 1 | Magnetic resonance imaging (MRI) in a scanning tunneling microscope (STM). a,** STM tip with a magnetic tip apex is scanned across a two-atom-thick layer of MgO on Ag(001) with single-atom spins deposited atop ($B_x = 0.9$ T, $B_z = 0.1$ T, $T = 1.2$ K). A radiofrequency (RF) voltage $V_{RF}$ at frequency $f$ is applied to the tunnel junction. During the scan, the magnetic tip induces a field gradient leading to an effective sweep in magnetic field $B_{tip}$ at the position of the atom. Thus, the atom spin is brought into resonance when the total magnetic field $B + B_{tip}$ matches the resonance frequency. **b,** Constant-current topography of a single hydrogenated titanium atom on an oxygen binding site of MgO (Ti$_O$). **c,** ESR spectrum taken on the center of the atom in (b) obtained by sweeping the frequency $f$ of the RF voltage and measuring the change in tunnel current $\Delta I$. Black line is a fitted Fano-Lorentzian. ($I = 10$ pA, $V_{DC} = 40$ mV, $V_{RF} = 60$ mV). **d,** MRI of the atom in (b), showing the spatially resolved spin resonance signal $\Delta I(x, y)$ for a fixed frequency $f = 25$ GHz. Light crescent shows the resonant slice, the set of tip positions where the Ti atom is in resonance: $f = f_0$ ($I = 35$ pA, $V_{DC} = 40$ mV, $V_{RF} = 25$ mV). **e,** Resonant slice imaged at different distances $z$ between tip and sample (Supplementary video 1, $I = 5 - 70$ pA, $V_{DC} = 40$ mV, $V_{RF} = 40 - 45$ mV, $f = 23.65$ GHz). $z = 0$ refers to point contact between tip and sample atom. The vertical images to the sides are cross sections centered on the atom. The topography of the Ti$_O$ atom is shown at the bottom.

Here, we combine an STM with ESR[13] to perform MRI scans on single-atom spin centers. In this approach, the magnetic-field gradient, the electric readout as well as the driving field are all combined in the STM tip apex. In contrast to existing scanning-field-gradient methods[21], our technique presently requires cryogenic temperatures, ultra-high vacuum, and placement of spin-resonant atoms near a conducting surface. Nevertheless, it allows MRI scans with sub-Ångström resolution. We additionally focus on the 3D magnetic interaction potential between tip spin and surface atom spin, which we derived from the MRI data. For different tip configurations and atomic species on the surface, we find fingerprints in the MRI scans that reflect the magnetic properties of both tip and surface atoms.

Figure 1a illustrates the experimental setup, consisting of single adatoms adsorbed on MgO and a magnetic STM tip. Figure 1b shows several adatoms on the surface, the central one being a hydrogenated Ti atom located on an oxygen binding site ($Ti_O$)[25]. Tunnel-current-detected ESR[13,14,26] performed by placing the STM tip above the atom allows its magnetic properties to be probed electrically. Here, a radiofrequency (RF) voltage $V_{RF}$ induces spin transitions between the ground and excited state, which are split by the Zeeman energy. This leads to a peak in tunnel current $\Delta I$ when the frequency $f$ of $V_{RF}$ is close to the atom's resonance frequency $f_0$ (Fig. 1c). The proximity of a magnetic STM tip applies a local magnetic field $\vec{B}_{tip}$ to the adatom, which shifts its Zeeman energy and leads to spatial variations in the resonance frequency $f_0$[25]. Consequently, the spatial pattern where resonance occurs can be mapped by scanning the tip across the atom with constant frequency $f$ (Supplementary Section 1). We scan the STM tip laterally in $x$ and $y$ while the vertical tip-atom distance $z$ is set by the tunnel conditions, making the tip follow the topographic height image acquired simultaneously (Fig. 1b). In Fig. 1d, the MRI scan shows a crescent-shaped resonance pattern for the tip near the atom. This "resonant slice"[8] depends on the vertical tip-sample distance $z$. For closer tip-surface distance, established by increasing the tunnel current setpoint $I$, the field $\vec{B}_{tip}$ present at the position of the atom changes, and as a result different resonant slice patterns are obtained (Fig. 1e).

The range of the MRI map in Fig. 1e spans ~1 Å vertically. The magnetic field gradient in this range is approximately $\Delta B_{\text{tip}}/\Delta z \approx 0.3\,\text{T/Å}$ (see Fig. S3) which is 4−5 orders of magnitude larger than in other scanning-field-gradient experiments[5,20]. This is a consequence of the close proximity between the magnetic tip and the surface atom, which are only a few Ångströms apart from point contact[27,28].

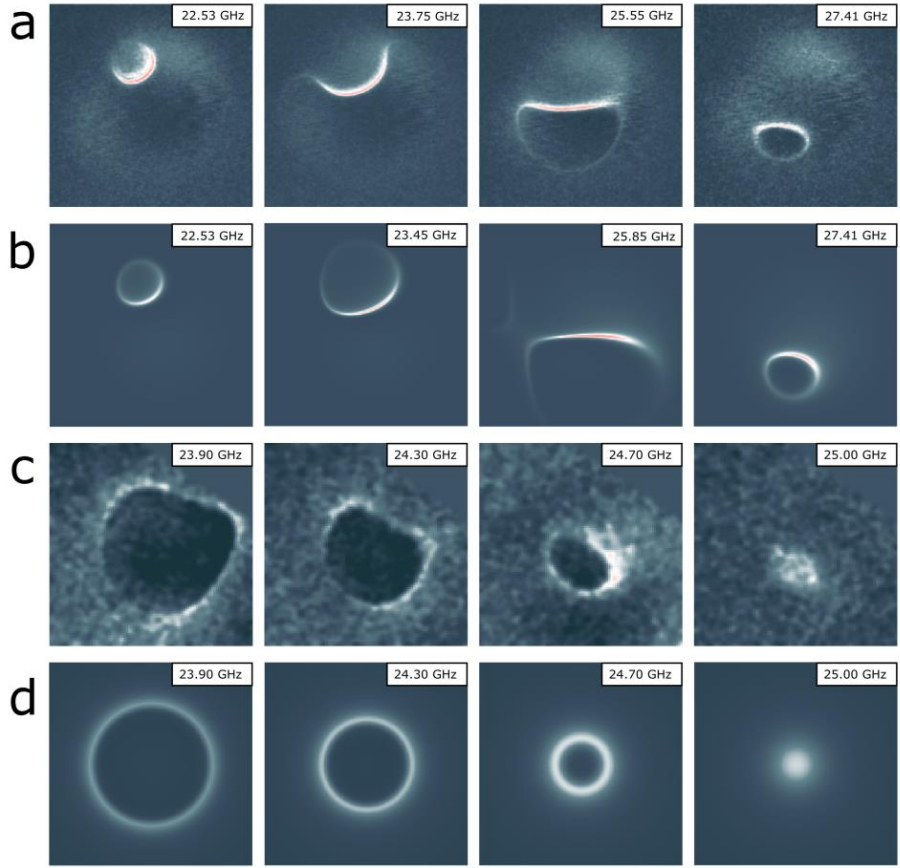

**Figure 2 | Mapping different magnetic interactions between tip and surface atom. a,** Resonant slices (light areas) of the Ti$_\text{O}$ atom in Fig. 1 at different frequencies $f$ [tunnel conditions: $I = 35\,\text{pA}$, $V_{\text{DC}} = 40\,\text{mV}$ (2.4 Å above point contact), $V_{\text{RF}} = 25\,\text{mV}$, image size: 15 Å × 15 Å]. **b,** Simulation of the resonant slices at similar energies as in (a) assuming magnetic dipole interaction $\vec{B}_{\text{dipole}}$ between tip and Ti atom (Supplementary Section 4). **c,** Resonant slices taken on another Ti$_\text{O}$ atom using a different magnetic tip [$I = 13\,\text{pA}$, $V_{\text{DC}} = 40\,\text{mV}$ (2.9 Å above point contact), $V_{\text{RF}} = 60\,\text{mV}$, image size: 13 Å × 13 Å]. **d,** Simulation of the resonant slices using exchange interaction $\vec{B}_{\text{exchange}}$ between tip and sample (Supplementary Section 4). Supplementary videos 2−5 show the continuous evolution of (a)−(d), respectively.

The resonant slice pattern is caused by the contribution of $\vec{B}_{tip}$ to the Zeeman energy of the surface atom. The resonant frequency at each position can be expressed by[13,25]

$$hf_0(x,y,z) = 2g\mu_B[\vec{B} + \vec{B}_{tip}(x,y,z)] \cdot \vec{S} \qquad (1)$$

where $h$ is Planck's constant, $\mu_B$ is the Bohr magneton and $g$ is the effective electron g-factor of the surface atom spin. $\vec{B}$ is the externally applied magnetic field and $\vec{S}$ the spin operator of the surface atom. For a fixed driving frequency $f$ the spatially-dependent tip field $\vec{B}_{tip}(x,y,z)$ can be used to tune the surface spin system into resonance. Depending on the tip magnetic field, the resonance condition will occur at a set of spatial positions that constitute the resonant slice. When varying the driving frequency $f$ at constant tunnel conditions the resonant slice changes as well, as shown in Fig. 2a for the same atom and tip as Fig. 1. Below, we use such a series of MRI scans acquired at different frequencies $f$ ($xyf$-measurements) to reveal the resonance frequency $f_0$, and thus the interaction potential, for each tip position in the images.

In our setup, magnetic tips are obtained by transferring several Fe atoms (typically 1–5) from the surface onto the apex of a non-magnetic metal tip. The MRI scans allow us to extract the magnetic properties of the tip by modeling its interaction with the surface atom spin. We use a simple model (Supplementary Section 4) to fit the resonant slices (Fig. 2b), that considers the relative alignment of one magnetic moment for the tip, and one for the surface atom, and their magnetic interaction according to Eq. (1). Here, we use magnetic dipolar interaction $\vec{B}_{tip} = \vec{B}_{dipole} = \frac{\mu_0 g_{tip}\mu_B}{4\pi|\vec{r}|^3} \cdot [\vec{S}_{tip} - 3(\vec{S}_{tip} \cdot \hat{r})\hat{r}]$, where $g_{tip}$, $\vec{S}_{tip}$ and $\hat{r}$ are the effective g-factor of the tip, the tip spin operator, and the unit vector from tip to surface atom, respectively. This describes well the main features of the resonant slices in Fig. 2a.

We find that the resonant slices change drastically for different magnetic tips. Figure 2c shows MRI scans using a different tip, revealing a nearly-circular resonant slice pattern. We attribute this to spatially isotropic exchange interaction[25,29] given by $\vec{B}_{tip} = \vec{B}_{exchange} = [J(\vec{r})/g\mu_B]\vec{S}_{tip}$ where $J(\vec{r})$ is the radial-dependent coupling constant. The respective simulations are shown in Fig. 2d.

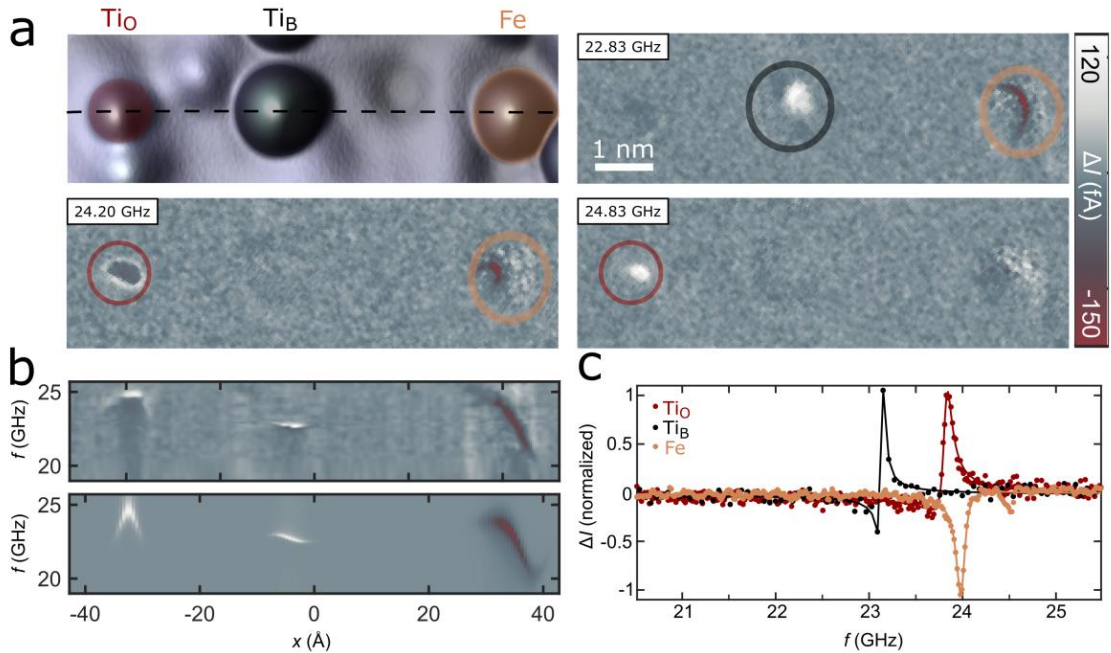

**Figure 3 | MRI of different atomic species**. **a,** Upper left: Constant-current STM image showing a hydrogenated Ti atom on an oxygen binding site (Ti$_O$, left), a hydrogenated Ti atom on a bridge binding site (Ti$_B$, middle) and an Fe atom (right) [$I = 10$ pA, $V_{DC} = 40$ mV, image size: 84 Å × 25 Å]. Remaining panels: $xy$-plane MRI scans of the same area for three different driving frequencies $f$ (Supplementary video 6). Colored circles highlight the topographic positions of the atoms. **b,** Upper part: MRI scan in the $xf$-plane along the dashed line in (a) [$I = 10$ pA, $V_{DC} = 40$ mV, $V_{RF} = 55$ mV]. Lower part: Simulations of the resonant slice pattern of the three atoms. For Ti$_B$ and Fe, magnetic dipolar interaction was used; for Ti$_O$ exchange interaction (Simulation details in Supplementary Section 4). **c,** ESR spectra taken on top of each individual atom ($I = 10$ pA, $V_{DC} = 40$ mV, $V_{RF} = 55$ mV). The peak intensity was normalized to ±1 due to different signal intensities for each atom. The sign of the ESR peak results from the relative alignment of tip and surface atom spins leading to a decrease in the spin-polarized tunneling contribution in the case of Fe[13].

Importantly, for any given STM tip, the resonant slice pattern also changes for different types of atoms on the surface. Figure 3a shows a topography of several magnetic atoms, a Ti$_O$ atom, a hydrogenated Ti atom on a bridge binding site (Ti$_B$) and an Fe atom (left to right), that were positioned into a line using atom manipulation. MRI scans at various frequencies reveal a unique resonant slice pattern for each atom. Additionally, both a resonant slice cross section (Fig. 3b) and ESR point spectra (Fig. 3c) highlight the change in interaction with the magnetic tip for each type of surface atom, by revealing differences in the frequencies, line broadening, and sign of the resonance signal. The different shapes of the resonant slices result from the different orientations of the atom spins on the surface as well as the respective interaction mechanism with the tip (Supplementary Section 4). The magnetic moment of Fe ($\mu_{Fe} = 5.4\,\mu_B$) is oriented out-of-plane due to magnetic anisotropy[30] while the spin of both Ti species ($\mu_{TiO} \approx \mu_{TiB} \approx 1\,\mu_B$) follows, to good approximation, the magnetic-field direction (Fig. 1a)[25,28]. Taking these spin properties of the surface atom spins into account and fitting the tip spin moment and direction, we are able to reproduce the resonant slice pattern (Fig. 3b). While in general each atomic species can couple both by exchange and by magnetic dipole coupling to the tip (see Fig. 2), we observe a tendency towards a preferred coupling mechanism, presumably caused by their difference in spin direction, topographic height and occupation of electronic orbitals (Supplementary Section 4 and 5).

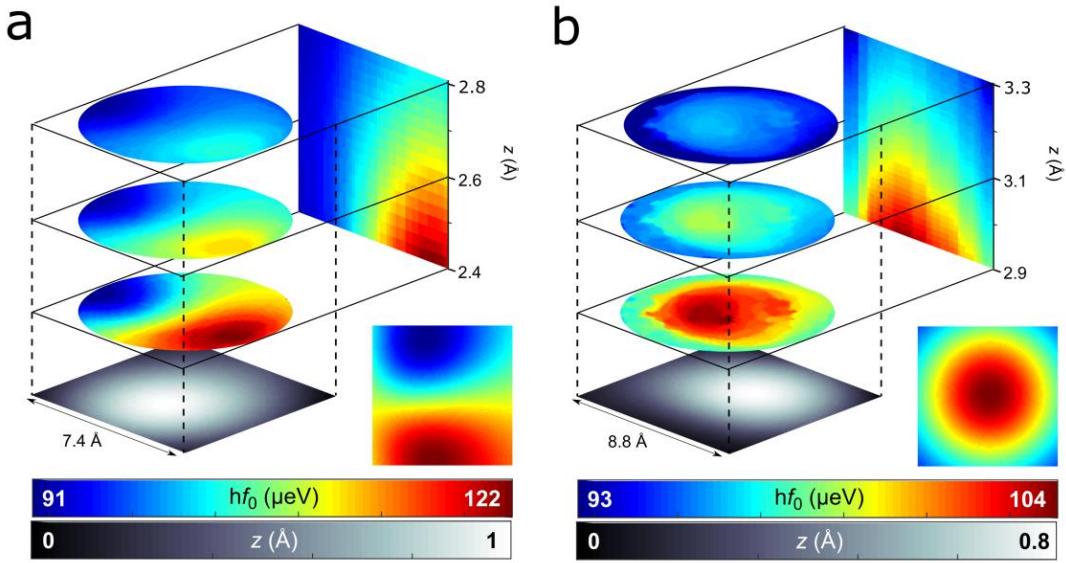

**Figure 4 | Mapping the 3D magnetic interaction potential. a,** Dipolar interaction potential $hf_0(x,y,z)$ derived from the resonant slices of the Ti$_O$ atom shown in Fig. 2a and additional data (Supplementary Section 3, $I = 8 - 55$ pA, $V_{DC} = 40$ mV, $V_{RF} = 12.5 - 40$ mV). Here, we plot the frequency $f$ at which $\Delta I$ is maximized for each point in space. The vertical image to the side is a cross section centered on the atom. The data were interpolated between discrete measurements of vertical distance $z$ to create a smooth image. The topography of the Ti$_O$ atom is shown at the bottom. The interaction potential reveals two poles, a unique feature of magnetic dipole interaction. Lower Inset: Magnetic dipole potential used for the simulations shown in Fig. 2b ($z = 2.4$ Å, Supplementary Section 4). **b,** Exchange interaction potential $hf_0(x,y,z)$ derived from the resonant slices of the Ti$_O$ atom shown in Fig. 2c and additional data (Supplementary Section 3, $I = 13$ pA, $V_{DC} = 40$ mV, $V_{RF} = 60$ mV). Lower Inset: Exchange potential used for the simulations in Fig. 2d ($z = 2.9$ Å Supplementary Section 4). $z = 0$ refers to point contact between tip and sample atom in (a) and (b).

The diversity of observed resonant slice patterns results from variations in orientation and magnitude of the tip spin (Fig. 2) and of the surface atom spins (Fig. 3). In contrast, atoms of the same element at the same type of binding site yield nearly identical resonance slices (Supplementary Section 2). In addition, we observe spin-spin interaction[30] and different states of the magnetic tip in the MRI scans (Supplementary Section 2).

For any given tip apex and surface atom, we can determine the resonance frequency $f_0(x, y)$ from a series of $xyf$-measurements (Supplementary Section 3). This is repeated for a series of tip heights $z$ to yield the full 3D interaction potential $hf_0(x, y, z)$ shown in Fig. 4 for the two atoms in Figs. 2a and 2c. These interaction energy maps show marked differences between the two tip apexes of Figs. 4a and 4b, most prominently the change in symmetry of the dominant interaction. This interaction symmetry is employed in simulations using magnetic dipolar and exchange potentials seen in the insets. Interestingly, by simulating the magnetic interaction potential of Eq. (1) we find that quantitative agreement cannot be obtained using a point-like spin on the tip. Instead, the spin needs to be distributed over a scale of a few Ångströms to adequately describe the data (Supplementary Section 4). We attribute this to the polarization of the surrounding tip metal atoms and the presence of multiple Fe atoms on the tip[31,32]. In general, the variety of magnetic interaction potentials found here (Supplementary Section 5) offers a range of possibilities in future experiments, for example to utilize the horizontal and vertical position of the tip to engineer the interaction given by Eq. (1) in order to controllably shift the energy levels of molecules and nanostructures.

Mapping the magnetic interaction between tip and surface atom gives access to the magnetic properties of both spin centers. Using a well characterized tip as demonstrated here enables the determination of the spin magnitude and orientation in unidentified spin structures by ESR and MRI. At present, MRI-STM already has the ability to reveal the unique spin signature of individual atomic species and locate their position and energy even in large-scale images (Supplementary Fig. S6), making the magnetic properties of large planar molecules and spin structures directly accessible.

## End Notes

**Acknowledgment.** We thank Bruce Melior for expert technical assistance. We gratefully acknowledge financial support from the Office of Naval Research. P.W., Y.J.B. and A.J.H. acknowledge support from the Institute for Basic Science under grant IBS-R027-D1. P.W. acknowledges support from the Alexander von Humboldt Foundation.

**Author contributions.** P.W. and C.P.L. conceived the experiment. P.W., K.Y. and Y.J.B. carried out the measurements. P.W. analyzed the data and wrote the manuscript. C.P.L and A.J.H supervised the project. All authors discussed the results and contributed to the manuscript.

**Additional information.** Supplementary information including videos of the MRI scans is available in the online version of the paper. Correspondence and requests for materials should be addressed to C.P.L. and A.J.H.

**Competing financial interests.** The authors declare no competing financial interests.